\documentclass[aps,prb,twocolumn,superscriptaddress,print]{revtex4-2}
\usepackage{amssymb,amsmath}
\usepackage{makecell}
\usepackage{rotating}
\usepackage{graphicx}
\usepackage{multirow}
\usepackage{hyperref}
\usepackage{appendix}
\usepackage{mathtools}
\usepackage{verbatim}
\usepackage{gensymb}
\usepackage{subfigure}
\usepackage{bm}
\usepackage{threeparttable}
\usepackage{booktabs}
\usepackage{braket}
\usepackage{xcolor}
\usepackage[normalem]{ulem}
\usepackage[mathscr]{euscript}
\usepackage{tabulary}
\usepackage{ragged2e}
\usepackage{adjustbox}
\usepackage{tabularx}
\usepackage{enumitem}
\setlist{nolistsep}

\setlist{
  noitemsep,
  topsep=0pt,    
  partopsep=0pt,  
  parsep=0pt,     
  itemsep=0pt      
}

\begin{document}
\title{Plasmons in bilayer graphene with O-type Kekul\'{e} distortion: Electron-hole asymmetry of the effective interlayer tunneling}
\author{Mingyu Wang}
\affiliation{Center for Quantum Sciences and School of Physics, Northeast Normal University, Changchun 130024, China}
\author{Guodong Yu}
\email{yugd000@nenu.edu.cn}
\affiliation{Center for Quantum Sciences and School of Physics, Northeast Normal University, Changchun 130024, China}
\author{Shengjun Yuan}
\affiliation{Key Laboratory of Artificial Micro- and Nano-structures of Ministry of Education and School of Physics, Wuhan University, Wuhan 430072, China}
\affiliation{Wuhan Institute of Quantum Technology, Wuhan 430206, China}
\affiliation{School of Artificial Intelligence, Wuhan University, Wuhan 430072, China}

\begin{abstract}
The effective interlayer tunneling (EIT) is a key quantity in determining the out-of-phase plasmons of a bilayer structure. Using the tight-binding model and the random phase approximation, we investigate plasmons in bilayer graphene with O-type Kekul\'{e} distortion. Both O/O and G/O bilayers exhibit strong electron-hole asymmetry in the EIT. For the O/O bilayer, the gap of the out-of-phase plasmon remains almost constant as the electron concentration varies, whereas it increases rapidly with increasing hole concentration; hole doping can also lead to stronger anomalous dispersion than electron doping. For the G/O bilayer, enhancing the Kekul\'{e} distortion $\Delta_0$ can weaken the EIT, as reflected by the reduced interlayer hybridization caused by the lifted layer degeneracy, especially for the electron-doped case. For the electron-doped G/O bilayer, the EIT is almost turned off when $\Delta_0=0.3$ from the perspective of the dispersion. Our results imply that switching the doping type between electron and hole can tune the performance of out-of-phase plasmon-based devices by modulating the EIT.
\end{abstract}

\maketitle

\section{Introduction}
Charge oscillation is one of the main low-energy collective excitations in metals and doped semiconductors\cite{Mahan2000Many,linhard}. It plays a crucial role in determining the optical and electronic properties of materials\cite{Pitarke_2007_RPP,GPlasmon_opti_natpho_2012,GPlasmon_opti_acsphot_2014}. For a coupled bilayer of two-dimensional electron gas (2DEG) in which the layers are coupled by the Coulomb interaction, there are two types of interlayer charge oscillations, namely the in-phase and out-of-phase modes\cite{bilayer_PRB_1981,bilayer_2DEG_DasSarma,bilayer_2DEG_asymmetry_PRB,Nlayer_plasmon}. The corresponding quasiparticles are called the in-phase and out-of-phase plasmons. For the bilayer 2DEG with the parabolic band structure, the dispersions of these two plasmon modes follow $\omega\sim\sqrt{q}$ and $\omega\sim q$, respectively, in the long-wavelength limit\cite{bilayer_PRB_1981}. In general, the frequency of the in-phase plasmon, which is associated with a positive interlayer Coulomb interaction, is higher than that of the out-of-phase plasmon, which is associated with a negative interlayer Coulomb interaction, because the negative interlayer Coulomb interaction reduces the potential energy and thus the restoring force during harmonic oscillation.

When interlayer tunneling is present, it leads to the formation of bonding and antibonding bands due to interlayer hybridization, and the dispersion of the out-of-phase plasmon deviates strongly from the $\omega\sim q$ behavior\cite{tunneling_prb_1995, bilayer_2DEG_DasSarma}. The most important feature is the appearance of a nonzero frequency gap in the long-wavelength limit. Since the bonding and antibonding states, as eigenstates of the bilayer, have equal projections onto the two layers, the out-of-phase charge oscillation can be activated solely by the coherence between the bonding and antibonding states. It is the main reason for the generation of the nonzero frequency gap of the out-of-phase plasmon at $q\rightarrow0$\cite{bilayer_2DEG_DasSarma}. Another important phenomenon is the anomalous dispersion of the out-of-phase plasmon at small momenta. The competition between the anomalous contribution mainly from the interlayer tunneling and the normal contribution from the Coulomb interaction results in a critical momentum that separates the anomalous and normal dispersion regions\cite{bilayer_2DEG_DasSarma}. 

\begin{figure*}[!htbp]
\centering
\includegraphics[width=0.7\textwidth]{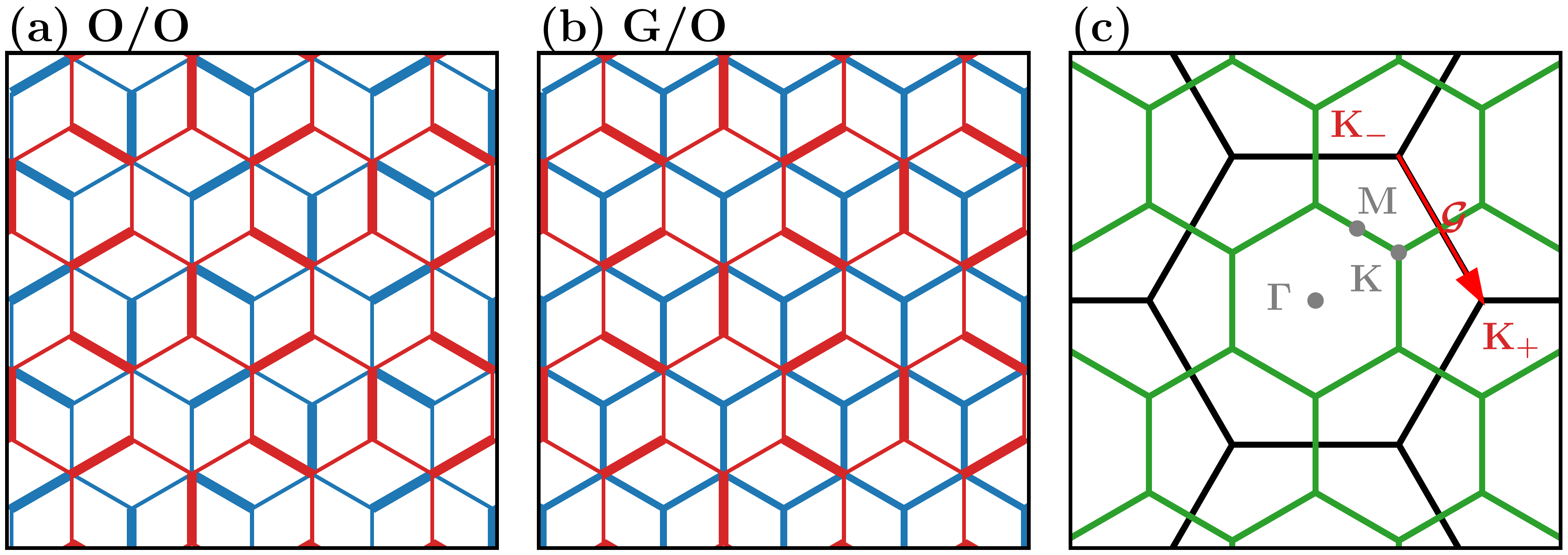}
\caption{(a) and (b) Schematic structures of the O/O and G/O bilayers. For the Kek-O layer, the thick and thin bonds represent the hopping amplitudes $-\gamma_0(1+2\Delta_0)$ and $-\gamma_0(1-\Delta_0)$, respectively. For the undistorted graphene layer, the nearest-neighbor hopping is $-\gamma_0$. (c) Brillouin zone (BZ) of the bilayer with Kekul\'{e} distortion bilayer, compared with that of undistorted graphene (black).} 
\label{fig:struct}
\end{figure*}

Before the discovery of graphene, plasmon studies in bilayer structures were generally carried out in double-quantum-well systems\cite{bilayer_well0,bilayer_well1,bilayer_well2,bilayer_well3}. With the explosive discovery and realization of 2D materials\cite{2D_materials0,2D_materials1,2D_materials2,2D_materials3,2D_materials4}, bilayers of 2D materials have become another platform for studying plasmons in bilayer structures\cite{plasmon_MoS2_bilayer,plasmon_blackP_bilayer,plasmon_MoS2_bilayer_AK,plasmon_bilayer_cuprate}. Among experimentally realized 2D materials, graphene is the most extensively studied\cite{plasmon_biG_MacDonald_PRB_2009, plasmon_DLG_EHH_2009, plasmon_AA_G_2013, plasmon_AA_G_2021, plasmon_bilayer_G, plasmon_biG_TonyLow, plasmon_tBG_nature, Plasmon_tBG_Tobias, plasmon_tBG_Kuang}. In experiments, graphene sheets are often grown on substrates\cite{graphene_growth_ACR_2013,graphene_growth_MT_2012,graphene_growth_small_2025}, which can also affect the electronic structure of graphene and modify the dielectric environment\cite{Gon_substrate_csr_2017,GonSiC_prl_2007,Gon_substrate_NM_2007}. With a suitable choice of substrate, Kekul\'{e} distortions of Y and O types (denoted Kek-Y and Kek-O) can form in graphene\cite{KekY_on_Cu,Kek_on_SiOx}. In addition, intercalation between graphene and the substrate is an effective method for introducing Kekul\'{e} distortion, such as Li, Eu, and Sn intercalation\cite{Kekule_Li_intercalation,Kekule_Li_intercalation_PRB,Kekule_Eu_intercalation,Kek_Sn_intercat,Kek_Sn_intercat_carbon}. Besides substrates and intercalation, Kekul\'{e} distortion can also be induced by periodic deposition of adatoms on graphene, such as Xe adatoms\cite{Kek_Xe_adatom}.

The Kekul\'{e} distortion can induce coupling between the two non-equivalent Dirac cones\cite{band_kekule,kekule_strain,KekOY_prb2020_transport,KekOY_prb2022_Klein,KekY_prb2022_zitterbewegung}. For Kek-Y distortion, this coupling does not break the linear dispersion but instead leads to two new Dirac cones with different Fermi velocities\cite{band_kekule,kekule_strain,KekOY_prb2020_transport,KekOY_prb2022_Klein,KekY_prb2022_zitterbewegung}. Owing to the similarity between the band structures of graphene and Kek-Y graphene, their plasmons are similar, especially at small electron concentrations. Although Kek-Y distortion introduces a new plasmon branch, it is highly unstable and rapidly damped by electron-hole excitations\cite{plasmon_kekY}. In contrast, for Kek-O distortion, the coupling between the two Dirac cones opens a band gap at $\Gamma$ and changes the dispersion from linear to parabolic\cite{band_kekule, kekule_strain, KekOY_prb2020_transport, KekOY_prb2022_Klein}. Therefore, stacking two Kek-O graphene layers yields a bilayer with parabolic bands and finite interlayer tunneling.

Because the electronic structure is strongly modified and the substrate influences the dielectric environment, plasmons in a Kek-O graphene bilayer should be effectively modulated by the Kekul\'{e} distortion. Importantly, it is necessary to examine whether the ideal 2DEG model with constant interlayer tunneling applies to bilayers of real 2D materials. In this paper, we focus on the out-of-phase plasmons of graphene bilayers with O-type Kekul\'{e} distortion, including two types of bilayers, O/O and G/O, as shown in Fig. \ref{fig:struct}(a) and (b), where O denotes the Kek-O layer and G the undistorted graphene layer. Our results indicate that the interlayer tunneling, as a key quantity, exhibits strong electron-hole asymmetry, leading to different out-of-phase plasmons in electron- and hole-doped O/O and G/O bilayers. Thus, switching the doping type between electron and hole provides an effective way to modulate the performance of plasmon-based devices. Because real bilayer materials deviate from the ideal bilayer 2DEG, we refer to the interlayer tunneling as the effective interlayer tunneling (EIT) hereafter.

\section{Methods}

\subsection{Tight-binding model}
For a graphene bilayer without Kekul\'{e} distortion, the lattice vectors are $\bm{a}_1=a(-\sqrt{3}/2,1/2)$ and $\bm{a}_2=a(\sqrt{3}/2,1/2)$, where $a=2.46$ \AA\ is the lattice constant. For the bottom layer, the positions for the A and B sublattices in a unit cell are $\bm{\tau}_A^b=\bm{a}_1/3+\bm{a}_2/3$ and $\bm{\tau}_B^b=2\bm{a}_1/3+2\bm{a}_2/3$, while for the top layer they are $\bm{\tau}_A^t=\bm{\tau}_A^b+\bm{v}_s$ and $\bm{\tau}_B^t=\bm{\tau}_B^b+\bm{v}_s$, where $\bm{v}_s=(\bm{a}_1+\bm{a}_2)/3+h\bm{e}_z$ for the AB-stacked bilayer. Here, $h=3.35$ \AA\ is the interlayer distance and $\bm{e}_z$ is the unit vector along the $z$-axis. For a site at $\bm{r}_A^l$ belonging to the A sublattice of layer $l$ ($l=b$ and $t$ for the bottom and top layers, respectively), there are three nearest neighbors in the B sublattice at $\bm{r}_B^l=\bm{r}_A^l+\bm{s}_u$ ($u=1,2,3$).

Kekul\'{e} distortion can be introduced into layer $l$ by using a non-uniform nearest-neighbor hopping energy expressed as\cite{band_kekule,kekule_strain}
\begin{equation}
t(\bm{r}_A^l, u) = -\gamma_0\left( 1+ 2 \mathrm{Re} \left[ \Delta_0 e^{i(p\bm{K}_+ +q\bm{K}_-)\cdot\bm{s}_u}e^{i\bm{\mathcal{G}}\cdot\bm{r}_A^l} \right] \right)
\label{eq:hop_kekule}
\end{equation}
for the hopping pair between a site at $\bm{r}_A^l$ and one of its nearest-neighbor sites at $\bm{r}_A^l+\bm{s}_u$, where $\bm{\mathcal{G}}=\bm{K}_+ -\bm{K}_-$ is the Kekul\'{e} vector [see Fig. \ref{fig:struct}(c)]. For given integers $p$ and $q$, Eq. \eqref{eq:hop_kekule} generates oscillating hoppings between $-\gamma_0(1-\Delta_0)$ and $-\gamma_0(1+2\Delta_0)$, leading to bond-density waves characterized by the index
\begin{equation}
\nu = 1+q-p \pmod 3.
\end{equation}
$\nu=0$ and $\pm1$ correspond to Kek-O and Kek-Y textures. Such non-uniform nearest-neighbor hoppings result in a larger unit cell, i.e., the $\sqrt{3}\times\sqrt{3}$ supercell relative to pristine graphene, as well as coupling between the $K_+$ and $K_-$ valleys with a coupling strength characterized by $\Delta_0$.

The tight-binding Hamiltonian of a bilayer graphene with Kekul\'{e} distortion can be decomposed as $H = \sum_{l}H_l + U$, where $H_l$ is the intralayer Hamiltonian of layer $l$, and $U$ is the interlayer coupling Hamiltonian. They are
\begin{equation}
\begin{split}
H_l & = \sum_{\bm{r}_A^l}\sum_{u=1}^3 t(\bm{r}_A^l,u) C_{\bm{r}_A^l}^{\dagger} C_{\bm{r}_A^l+\bm{s}_u} + \mathrm{H.c.},\\
U   & = \sum_{\bm{r}_X^b,\bm{r}_{X^{\prime}}^t}\sum_{X,X^{\prime}}  t^{\prime}(\bm{r}_X^b-\bm{r}_{X^{\prime}}^t) C_{\bm{r}_X^b}^{\dagger} C_{\bm{r}_{X^{\prime}}^t} + \mathrm{H.c.},
\end{split}
\label{eq:H_real}
\end{equation}
where $C_{\bm{r}}$ and $C_{\bm{r}}^{\dagger}$ are the annihilation and creation operators at site $\bm{r}$. Here, $t^{\prime}(\bm{r}_X^b-\bm{r}_{X^{\prime}}^t)$ is the hopping energy between the $X$ sublattice of layer $b$ at $\bm{r}_X^b$ and the $X^{\prime}$ sublattice of layer $t$ at $\bm{r}_{X^{\prime}}^t$. It is given by\cite{LCAO}
\begin{equation}
t^{\prime}(d\bm{r}) = n^2 V_{pp\sigma}(|d\bm{r}|)+(1-n^2)V_{pp\pi}(|d\bm{r}|),
\label{eq:TB}
\end{equation}
where $n$ is the direction cosine of $d\bm{r}$ with respect to the $z$ axis, and the Slater-Koster parameters $V_{pp\sigma}$ and $V_{pp\pi}$ read
\begin{equation}
\begin{split}
V_{pp\pi}(|d\bm{r}|)&=-\gamma_0 e^{\delta(a_0-|d\bm{r}|)}, \\
V_{pp\sigma}(|d\bm{r}|)&=\gamma_1 e^{\delta(h_0-|d\bm{r}|)}.
\end{split}
\label{eq:TB_SK}
\end{equation}
In this paper, we use the following tight-binding parameters: $\gamma_0=2.8$ eV, $a_0=1.42$ \AA, $\gamma_1=0.48$ eV, $h_0=3.35$ \AA, and $\delta=2.218$ \AA$^{-1}$. Interlayer hoppings with a site-site distance larger than 5 \AA\ are neglected.

\subsection{Dielectric tensor}

To identify the plasmons in a bilayer structure, including both in-phase and out-of-phase modes, we compute the $2\times2$ dielectric tensor. Within the random phase approximation (RPA) \cite{rpa,jpsj_Lin,PRB_Lin,PLA_Lin}, it reads
\begin{equation}
\varepsilon(\bm{q},\omega) = \mathcal{I}_2 - V(\bm{q}) \Pi^{0}(\bm{q},\omega),
\label{eq:epsilon_tensor}
\end{equation}
where $\mathcal{I}_2$ denotes the $2\times2$ identity matrix, and
\begin{equation}
V(\bm{q}) = v_q \begin{pmatrix}
1 & e^{-q h} \\
e^{-q h} & 1
\end{pmatrix},
\label{eq:vq_tensor}
\end{equation}
with $v_q = \frac{e^2}{2\varepsilon_0 \varepsilon_r q}$. Here, $V(\bm{q})$ is the Coulomb potential tensor in momentum space, which depends on the dielectric environment $\varepsilon_r$, and $v_q$ and $v_q e^{-qh}$ are the intralayer and interlayer Coulomb potentials, respectively. The non-interacting dynamic polarization $\Pi^{0}(\bm{q},\omega)$ is evaluated via the Lindhard formula \cite{linhard,jpsj_Lin,PRB_Lin,PLA_Lin}:
\begin{equation}
\begin{split}
\Pi^0_{l l^{\prime}}&(\bm{q},\omega) \\
&= \frac{g_s}{4\pi^2} \sum_{n n^{\prime}} \int_{\text{BZ}} d\bm{k} 
\frac{n_F(E_{n,\bm{k}}) - n_F(E_{n^{\prime},\bm{k}+\bm{q}})}{E_{n,\bm{k}} - E_{n^{\prime},\bm{k}+\bm{q}} + \hbar\omega + i\eta} \\
&\quad \times \langle n^{\prime},\bm{k}+\bm{q} | P_l | n,\bm{k} \rangle  
\langle n,\bm{k} | P_{l^{\prime}} | n^{\prime},\bm{k}+\bm{q} \rangle,
\end{split}
\label{eq:Pi_tensor}
\end{equation}
where $g_s = 2$ accounts for spin degeneracy, $P_l$ is the layer projection operator that extracts the component of the wavefunction in layer $l$, and $\eta \to 0^+$ is a positive infinitesimal, with $\eta = 0.002$ eV adopted in this work. The BZ is sampled using a $1200\times 1200$ k-point mesh to obtain the converged plasmon dispersions with the convergence test in the supplementary material (SM). The eigenstates $|n,\bm{k}\rangle$ and eigenvalues $E_{n,\bm{k}}$ are obtained by diagonalizing the full tight-binding Hamiltonian, and $n_F$ is the Fermi--Dirac distribution function with the temperature set to 300 K in this paper. Using Eqs.~\eqref{eq:epsilon_tensor}, \eqref{eq:vq_tensor}, and \eqref{eq:Pi_tensor}, we obtain the dielectric tensor that incorporates both interlayer Coulomb coupling and the EIT.

After deriving the $2\times2$ dielectric matrix $\varepsilon$ at arbitrary $\bm{q}$ and $\omega$, we first diagonalize it to obtain its two eigenvalues $\varepsilon_1$ and $\varepsilon_2$. For each eigenvalue, the electronic energy loss function (EELF) can be defined as
\begin{equation}
\mathcal{L}(\bm{q}, \omega) = -\mathrm{Im}\left[\frac{1}{\varepsilon_{i}(\bm{q}, \omega)}\right].
\label{eq:eelf}
\end{equation}
For a given momentum $\bm{q}$, a plasmon with frequency $\omega_p$ can be identified if the EELF exhibits a sharp peak at $\omega_p$. The eigenvector associated with this eigenvalue $\varepsilon_i(\bm{q},\omega_p)$ then gives the phase relationship of the interlayer charge oscillation for this plasmon. If the dielectric tensor is symmetric, i.e., $\varepsilon_{11}=\varepsilon_{22}$ and $\varepsilon_{12}=\varepsilon_{21}$, as in an AB-stacked bilayer, the plasmons correspond purely to in-phase or out-of-phase oscillations between the layers, and the eigenvectors are always $\bm{u}_{in}=1/\sqrt{2}[1,1]$ and $\bm{u}_{out}=1/\sqrt{2}[1,-1]$. If the symmetry of the dielectric tensor is broken, for example by the application of Kekul\'{e} distortion, there can be the mixed phase plasmon besides the in-phase and out-of-phase plasmons.

Besides the peaks in the EELF, the plasmon frequency at a fixed $\bm{q}$ can also be obtained from the zeros of the real parts of the dielectric functions $\varepsilon_1(\omega,q)$ and $\varepsilon_2(\omega,q)$. In numerical calculations, finding the zeros of a function is easier than locating its peaks, especially for functions with background noise. After determining the plasmon frequency at $\bm{q}$, the lifetime can be derived as $1/(2\gamma)$, where
\begin{equation}
\gamma = \frac{\mathrm{Im}[\varepsilon_i(\bm{q},\omega_p)]}{\left.\frac{\partial \mathrm{Re}[\varepsilon_i(\bm{q},\omega)]}{\partial \omega}\right|_{\omega=\omega_p}}
\label{eq:gamma}
\end{equation}
Here, $\mathrm{Re}[\varepsilon_i]$ and $\mathrm{Im}[\varepsilon_i]$ are the real and imaginary parts of the dielectric function $\varepsilon_i$. A stable and observable plasmon should have a large EELF value and a long lifetime $\tau=\hbar/(2\gamma)$ value. To derive the plasmon dispersions, we consider the zeros of $\mathrm{Re}[\varepsilon_i(q,\omega_p)]=0$ for which the sign of $\mathrm{Re}[\varepsilon_i]$ changes from negative to positive as the frequency increases. To identify observable plasmons, we focus on those with $\mathcal{L}(\omega_p, q)>0.5$ and $\gamma(\omega_p,q)>0$. It is known that $\gamma$ depends on $\eta$, so an accurate $\gamma$ should be derived by extrapolation at $\eta\rightarrow0$. But we focus on its change trend with varying parameters when presenting the plasmon dispersions, so $\eta$ is fixed at 0.002 eV for all results shown in this paper.

\begin{figure}[!htbp]
\centering
\includegraphics[width=0.48\textwidth]{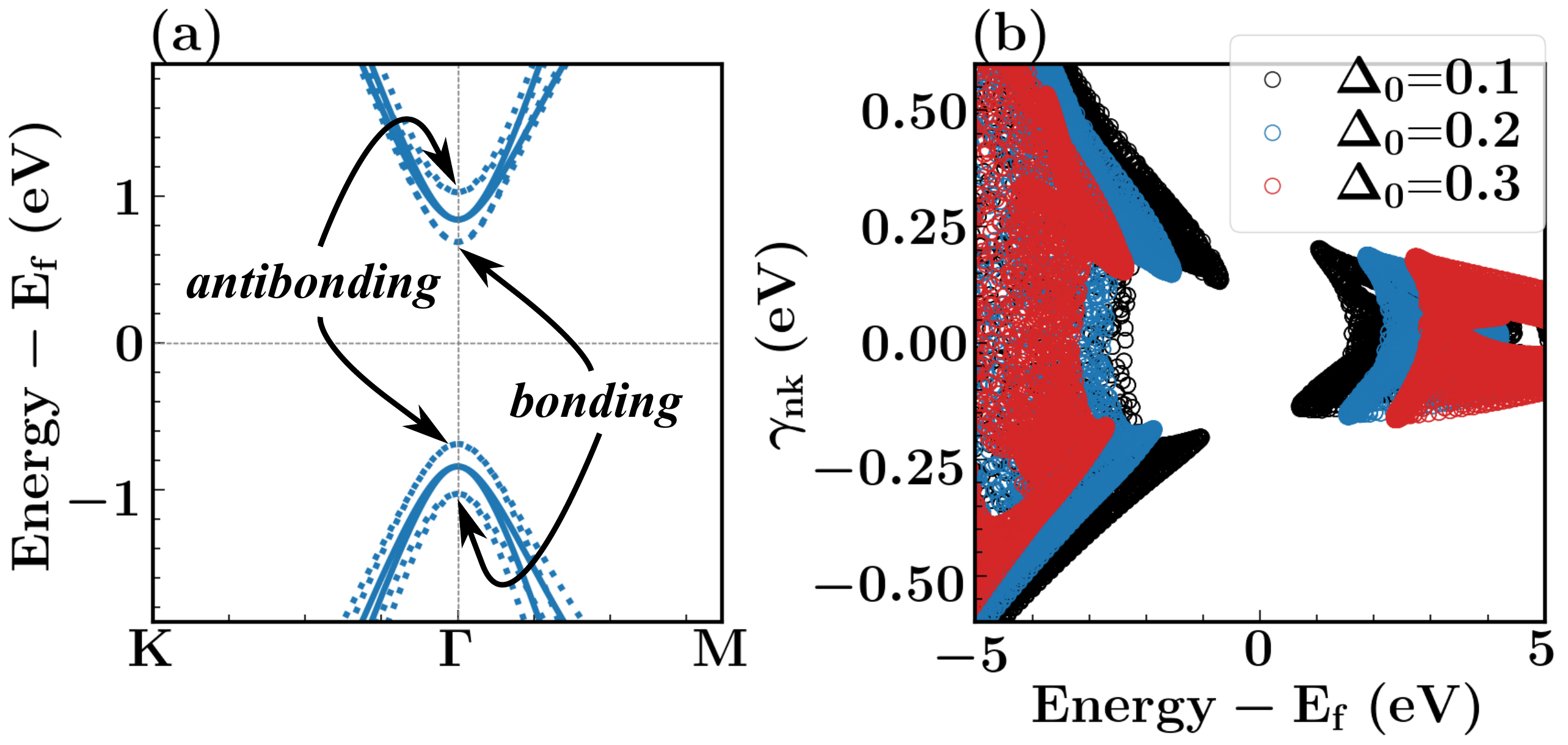}
\caption{(a) Band structure of the O/O bilayer with $\Delta_0=0.1$ plotted in dashed blue lines. The solid blue lines show the bands of the Kek-O monolayer with $\Delta_0=0.1$ for comparison. The bonding and antibonding bands are labeled. (b) The $\gamma_{nk}$ values of the O/O bilayer for several $\Delta_0$ values.} 
\label{fig:band_ihs_OO}
\end{figure}

\section{Results and discussion for O/O bilayer}

\subsection{Band structure}
For Kek-O graphene, the low-energy band structure is given by $E_{\pm}=\pm\sqrt{\hbar^2v_0^2k^2+(3\gamma_0\Delta_0)^2}$\cite{band_kekule}, and thus it has a band gap of $6\gamma_0\Delta_0$ at $\Gamma$, where $v_0=3\gamma_0 b/(2\hbar)$ is the Fermi velocity of pristine graphene and $b=a/\sqrt{3}$ is the carbon-carbon bond length. Around $\Gamma$, we expand the dispersion as $E_{\pm} \approx \pm (3\gamma_0\Delta_0 + \frac{1}{2}\frac{v_0^2}{3\gamma_0\Delta_0}\hbar^2k^2)$, which gives the effective mass
\begin{equation}
m^*=3\gamma_0\Delta_0/v_0^2
\label{eq:eff_mass_D0}
\end{equation}
for both the electron and hole bands. For $\Delta_0=0.1$, 0.2, and 0.3, we obtain $m^*=0.18m_e$, $0.36m_e$, and $0.54m_e$, respectively. Therefore, a stronger Kekul\'{e} distortion results in a larger effective mass.

As an example, Fig. \ref{fig:band_ihs_OO}(a) shows the band structures of the O/O bilayer with $\Delta_0=0.1$. Our results indicate that the two layers hybridize with each other, and each band of the Kek-O monolayer splits into a higher-energy antibonding band and a lower-energy bonding band. Consequently, the O/O bilayer exhibits a band gap smaller than $6\gamma_0\Delta_0$, because after hybridization its CBM and VBM originate from the bonding band on the electron side and the antibonding band on the hole side, respectively. Owing to the strong interlayer hybridization and the sizable energy difference between the bonding and antibonding states, the EIT is non-negligible and strongly affects the out-of-phase plasmon. In addition, the interlayer hybridization does not alter the effective mass of the Kek-O monolayer.

\begin{figure*}[!htbp]
\centering
\includegraphics[width=0.9\textwidth]{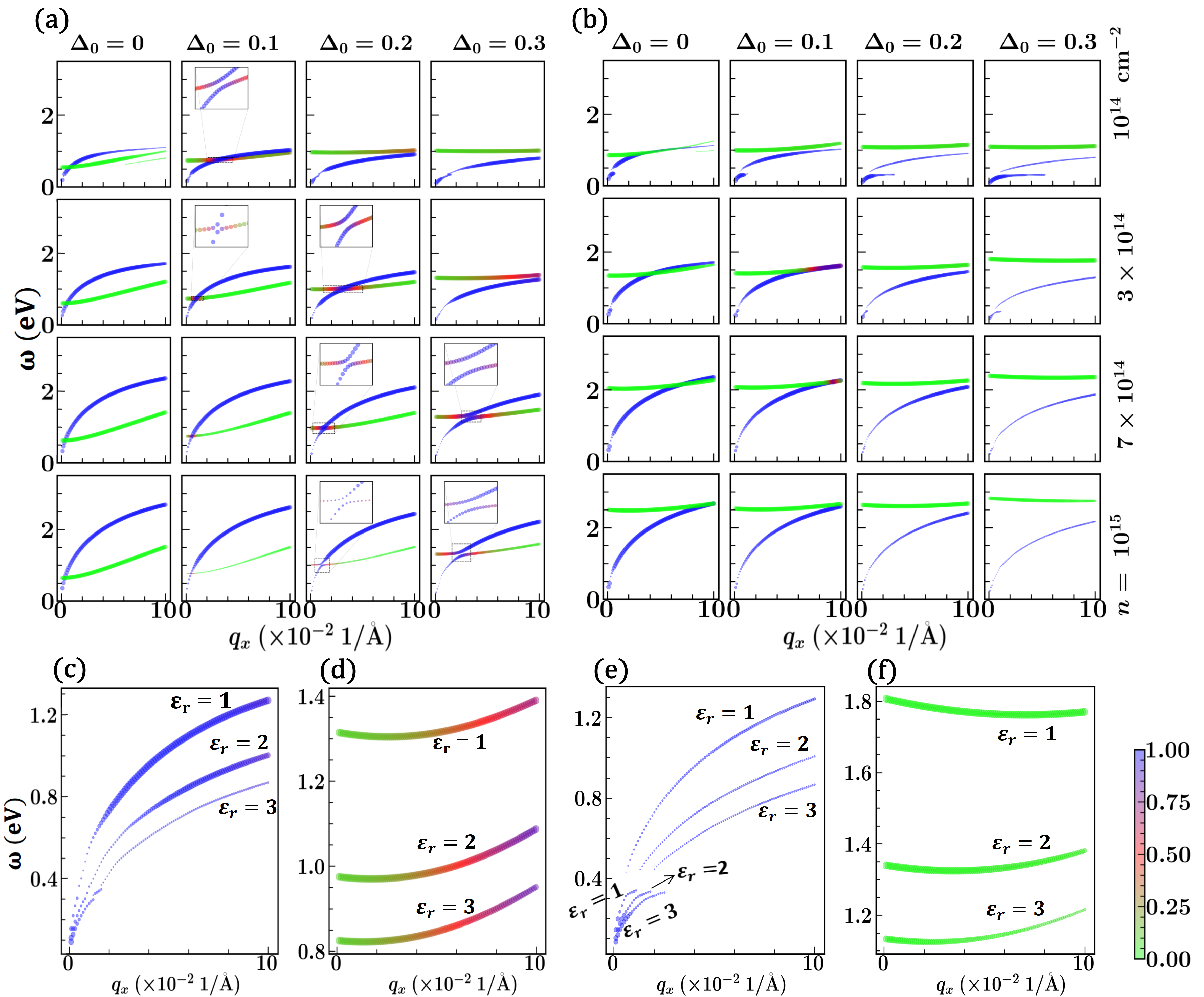}
\caption{(a) and (b) Plasmons of the electron- and hole-doped O/O bilayer. In (a) and (b), from top to bottom, the electron or hole concentration $n$ varies from $10^{14}$ to $10^{15}\ \mathrm{cm}^{-2}$ through $3\times10^{14}$ and $7\times10^{14}\ \mathrm{cm}^{-2}$; from left to right, $\Delta_0$ is $0, 0.1$, $0.2$, and $0.3$. For all calculations in this figure, the background dielectric constant is $\varepsilon_r=1$. The colorbar shows the projection of the eigenstates of the dielectric matrix onto the vector $[1,1]/\sqrt{2}$, so that blue (projection=1) and green (projection=0) correspond to the pure in-phase and out-of-phase plasmons, respectively, while other colors indicate mixed-phase plasmons. (c)-(f) Plasmons of the electron- and hole-doped O/O bilayer for several background dielectric constants with $\Delta_0=0.3$ and $n=3\times10^{14}\ \mathrm{cm}^{-2}$. Panels (c) and (d) correspond to the electron-doped bilayer, while (e) and (f) correspond to the hole-doped bilayer. Panels (c) and (e) show the in-phase plasmons, and (d) and (f) show the out-of-phase plasmons. The dot size in each subplot is proportional to $1/(2\gamma)$, so a dot with larger size corresponds to the larger lifetime and smaller Landau damping.} 
\label{fig:plasmon_OO}
\end{figure*}

\subsection{Plasmons}
In Fig. \ref{fig:plasmon_OO}, we show the plasmons of electron- and hole-doped O/O bilayers for $\Delta_0$ ranging from $0.0$ to $0.3$ and carrier concentration $n$ ranging from $10^{14}$ to $10^{15}\ \mathrm{cm}^{-2}$. One in-phase and one out-of-phase mode appear in the low-frequency region. Compared with the case $\Delta_0=0$, the Kekul\'{e} distortion can result in anti-crossing behavior, as shown in the insets of Fig. \ref{fig:plasmon_OO}(a). Specifically, when the in-phase and out-of-phase plasmons meet, they couple to each other and their dispersions are renormalized. Consequently, a gap opens around the crossing point, and the phase becomes mixed between in-phase and out-of-phase characters. As expected, the dispersions of both the in-phase and out-of-phase modes can be tuned by the carrier concentration $n$, the Kekul\'{e} distortion strength $\Delta_0$, and the background dielectric constant $\varepsilon_r$. In this paper, we present the plasmons for momentum $\bm{q}$ along only the $x$-axis (zigzag direction), because the plasmons for momentum $\bm{q}$ along the $y$-axis (armchair direction) give similar physics, as shown in the SM.

\subsubsection{In-phase plasmon}

Within classical electromagnetic theory, the in-phase charge oscillation can be understood as follows. We assume that the charge density forms a plane wave, with the displacement field given by $\bm{u}(\bm{r},t)=\bm{u}(\bm{q},\omega) e^{i(\bm{q}\cdot\bm{r}-\omega t)}$ and $\bm{u}\parallel\bm{q}$ for a longitudinal wave. The density distribution is written as $n(\bm{r},t)=n_0+\delta n(\bm{r},t)$, where $n_0$ is the density without oscillation and $\delta n(\bm{r},t)=\delta n(\bm{q},\omega)e^{i(\bm{q}\cdot\bm{r}-\omega t)}$ is the density fluctuation due to the oscillation. Keeping only the linear term, the current density field is obtained as $\bm{j}(\bm{r},t)=n_0 \frac{\partial \bm{u}}{\partial t}=-i\omega n_0\bm{u}(\bm{r},t)$. The continuity equation yields $\delta n(\bm{q},\omega)=-in_0 u(\bm{q},\omega)q$. The Coulomb potential energy is $\frac{1}{2}\delta n(\bm{q},\omega)[\phi(\bm{q},\omega)]^*=\frac{1}{2} n_0^2 q^2v_q u^2$. By analogy with a harmonic oscillator, we obtain the frequency $\omega^2=\frac{K_{\rm eff}}{n_0 m^*}=\frac{n_0^2 q^2v_q}{n_0 m^*}$, so the in-phase plasmon has the dispersion\cite{bilayer_2DEG_DasSarma}
\begin{equation}
\omega\sim \sqrt{\frac{n}{\varepsilon_r m^*}q}.
\end{equation}
Therefore, as shown in Fig. \ref{fig:plasmon_OO}(a), (b), (c), and (e), increasing the carrier concentration increases the slope of the dispersion; increasing the background dielectric constant reduces the Coulomb potential and thus lowers the frequency; and enhancing the Kekul\'{e} distortion increases the effective mass according to Eq. \eqref{eq:eff_mass_D0}, thereby reducing the slope of the dispersion. In addition, especially for the hole-doped bilayer, enhancing the Kekul\'{e} distortion can enhance Landau damping.

\subsubsection{The gap of the out-of-phase plasmon}

\begin{figure}[!htbp]
\centering
\includegraphics[width=0.46\textwidth]{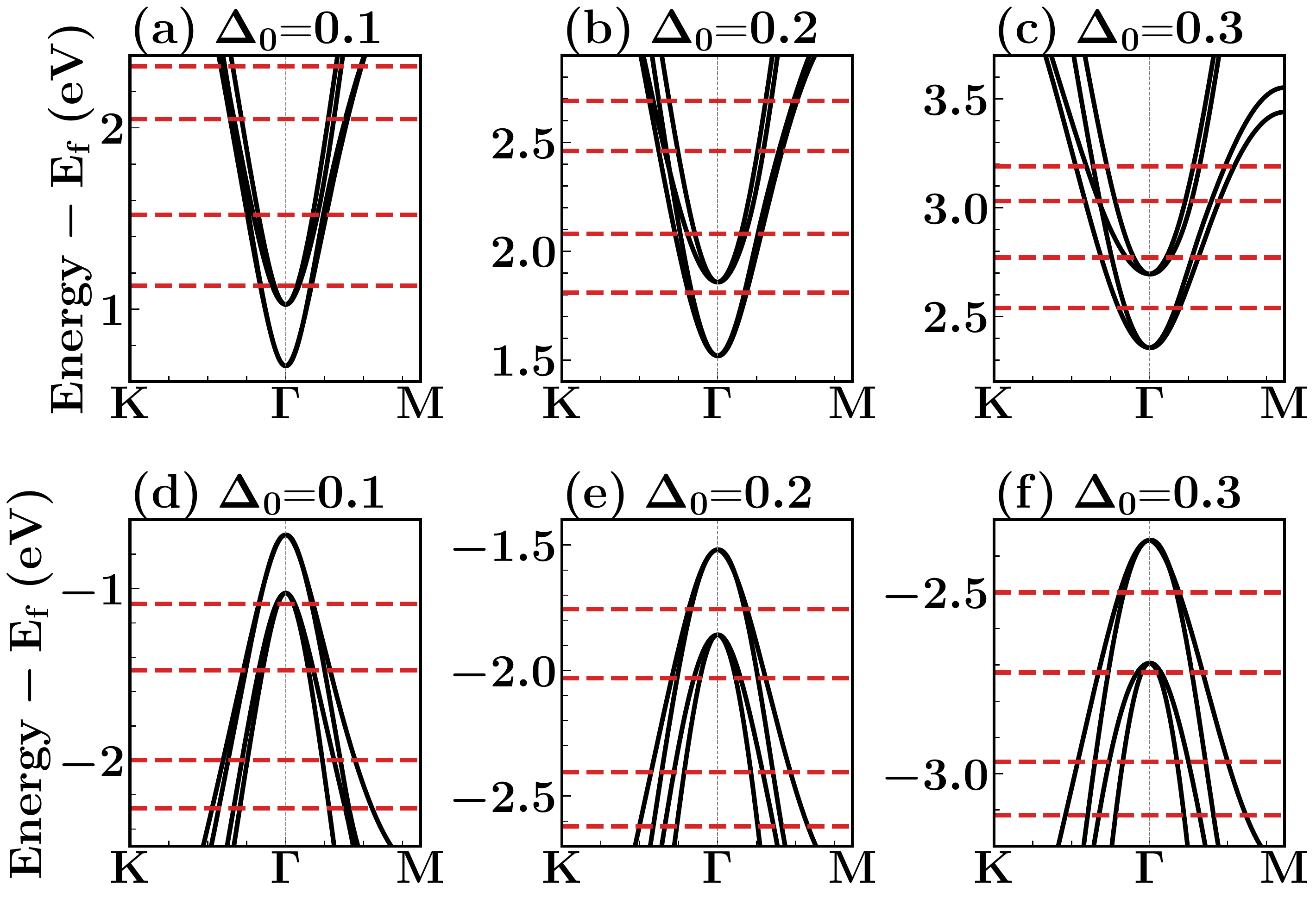}
\caption{Chemical potentials of the O/O bilayer with $\Delta_0 = 0.1, 0.2$, and $0.3$. In each subplot, the four red dashed lines indicate the chemical potentials corresponding to carrier concentrations of $10^{14}$, $3\times10^{14}$, $7\times10^{14}$, and $10^{15}\ \mathrm{cm}^{-2}$ for electron doping in (a)--(c) and hole doping in (d)--(f). A chemical potential closer to the Fermi energy corresponds to a smaller carrier concentration.} 
\label{fig:mus}
\end{figure}

For a coupled bilayer 2DEG, the out-of-phase plasmon gap, determined by the EIT and the interlayer Coulomb interaction, is given by\cite{bilayer_2DEG_DasSarma}
\begin{equation}
\Delta^2 = (\Delta_{SAS}^{\rm eff})^2 + \frac{\pi\hbar^2}{m^*}(n_+ - n_-)(q_{TF}h)\Delta_{SAS}^{\rm eff},
\label{eq:gap_original}
\end{equation}
where $\Delta_{SAS}^{\rm eff}$ is the EIT strength, a constant defined as the energy difference between the symmetric bonding and antisymmetric antibonding bands, $q_{TF}=2m^*e^2/(4\pi\varepsilon_0\varepsilon_r\hbar^2)$ is the 2D Thomas-Fermi wave vector, and $n_{\pm}=\bar{n}\pm n_c$ are the electron concentrations in the bonding and antibonding bands, respectively, with $\bar{n}=n/2$ being the average carrier concentration per layer and $n$ the total carrier concentration of the bilayer system. Here, $n_c=\frac{m^*}{2\pi \hbar^2}\Delta_{SAS}^{\rm eff}$ when both the bonding and antibonding bands are occupied, and $n_c=\bar{n}$ when only the bonding band is occupied. For most carrier concentrations considered here, both the bonding and antibonding bands are occupied, as shown in Fig. \ref{fig:mus}. Therefore, Eq. \eqref{eq:gap_original} becomes
\begin{equation}
\Delta=\Delta_{SAS}^{\rm eff}\sqrt{1+\frac{2m^*e^2h}{4\pi\varepsilon_0\varepsilon_r\hbar^2}}.
\label{eq:gap}
\end{equation}

As in the ideal bilayer 2DEG, the bonding and antibonding bands of the O/O bilayer with Kekul\'{e} distortion can be clearly identified due to interlayer hybridization, as shown in Fig. \ref{fig:band_ihs_OO}(a). However, unlike the ideal bilayer 2DEG, the EIT strength cannot be regarded as a constant over the entire energy range, because the energy difference between the bonding and antibonding bands is not a simple constant. Therefore, the plasmon behavior of the O/O bilayer can be understood using the ideal bilayer 2DEG model but with an energy- (or doping-) dependent EIT. For the O/O bilayer with $\Delta_0=0$, however, the parabolic bands arise from AB stacking rather than the Kekul\'{e} distortion, and the bonding and antibonding bands cannot be clearly identified. Consequently, Eqs. \eqref{eq:eff_mass_D0} and \eqref{eq:gap_original} are no longer applicable. Hence, we focus on O/O bilayers with finite Kekul\'{e} distortion in the following. Before addressing the EIT, we first examine the interlayer hybridization, defined as $\gamma_{nk}=\langle \Phi_{nk}|U|\Phi_{nk} \rangle$, where $U$ is the interlayer coupling Hamiltonian and $\Phi_{nk}$ is the eigenstate of the bilayer. For a given eigenstate, $\gamma_{nk}$ represents the average energy shift induced by interlayer hybridization and thus provides a qualitative measure of the EIT.

In Fig. \ref{fig:band_ihs_OO}(b), we plot the interlayer hybridization strength for the states of the O/O bilayer with $\Delta_0=0.1$, $0.2$, and $0.3$. The positive and negative $\gamma_{nk}$ correspond to the antibonding and bonding states, respectively. Our results indicate that the interlayer hybridization exhibits strong electron-hole asymmetry and is much more sensitive to hole doping than to electron doping. Note that $\gamma_{nk}$ is a local single-particle hybridization measure and is not identical to the EIT $\Delta_{SAS}^{\rm eff}$ that enters the long-wavelength plasmon gap. In the ideal 2DEG limit, the two coincide, but in a real bilayer the latter can be obtained from the out-of-phase plasmon gap via Eq. \eqref{eq:gap} by fitting with the bilayer 2DEG model. Solving Eq. \eqref{eq:gap} for the EIT yields $\Delta_{SAS}^{\rm eff} = \Delta [1 + 2m^*e^2h/(4\pi\varepsilon_0\varepsilon_r\hbar^2)]^{-1/2}$. The energy-dependent variation of $\gamma_{nk}$ does not necessarily translate into a variation of the plasmon gap, but the electron-hole asymmetry of the EIT can still be anticipated from the interlayer hybridization results, since both capture the same qualitative trend even though they are not identical in value.

In Fig. \ref{fig:plasmon_OO}(a), we show the plasmons of the electron-doped O/O bilayer. We focus on electron concentrations for which both the bonding and antibonding bands are occupied for all three $\Delta_0$ values, as shown in Fig. \ref{fig:mus}. This corresponds to the chemical potential lying above the antibonding band minimum on the electron side. These concentrations include $n=10^{14}$, $3\times10^{14}$, $7\times10^{14}$, and $10^{15}\ \mathrm{cm}^{-2}$ for $\Delta_0=0.1$, and $n=3\times10^{14}$, $7\times10^{14}$, and $10^{15}\ \mathrm{cm}^{-2}$ for $\Delta_0=0.2$ and 0.3. Our results indicate that the gap $\Delta$ remains constant as the electron concentration varies, and the three Kekul\'{e} distortion strengths $\Delta_0=0.1$, 0.2, and 0.3 yield gaps $\Delta$ of 0.74, 1.0, and 1.31 eV, respectively. In this work, the out-of-phase plasmon gap is extracted at a small momentum $q=0.001\ \mathrm{\AA}^{-1}$ to avoid the numerical divergence at $q=0$. Using Eq. \eqref{eq:gap}, we further obtain EIT strengths of $\Delta_{SAS}^{\rm eff}=0.41$, 0.42, and 0.47 eV for the three $\Delta_0$ values. These results indicate that, for the electron-doped O/O bilayer, the EIT is nearly constant with respect to both the electron concentration and the Kekul\'{e} distortion strength, although enhancing the Kekul\'{e} distortion increases the out-of-phase plasmon gap.

\begin{figure}[!htbp]
\centering
\includegraphics[width=0.35\textwidth]{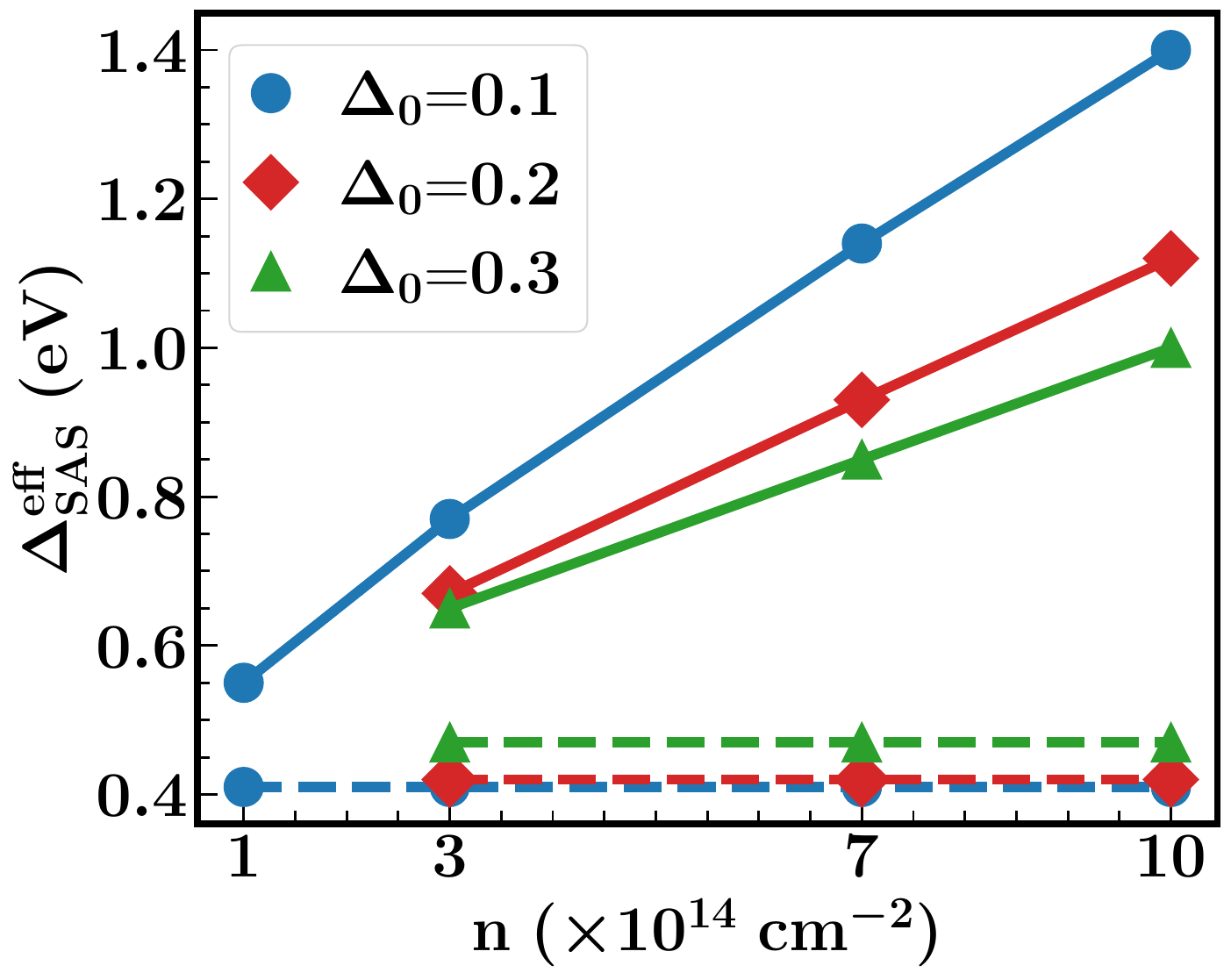}
\caption{The dependence of the EIT strength on the carrier concentration for O/O bilayer. The solid and dashed lines are for the hole- and electron-doped bilayers, respectively.} 
\label{fig:Dsas_hole}
\end{figure}

In Fig. \ref{fig:plasmon_OO}(b), we show the plasmons of the hole-doped O/O bilayer. For all three $\Delta_0$ values, as shown in Fig. \ref{fig:mus}, the same carrier concentration as in the electron-doped case leads to holes occupying both the bonding and antibonding bands, which means that the chemical potential lies below the bonding band maximum on the hole side. For this hole-doped case, $n_{\pm}$ in Eq. \eqref{eq:gap_original} denotes the hole concentrations in the antibonding and bonding bands on the hole side, respectively. Our results indicate that the out-of-phase plasmon gap is sensitive to the hole concentration for any $\Delta_0$. Using Eq. \eqref{eq:gap}, we derive the EIT strength $\Delta_{SAS}^{\rm eff}$ for the hole-doped O/O bilayer, as shown in Fig. \ref{fig:Dsas_hole}. Our results indicate that increasing the hole concentration enhances the EIT strength, and hole doping results in a stronger EIT than electron doping. In addition, enhancing the Kekul\'{e} distortion reduces the EIT, although it slightly increases the out-of-phase plasmon gap.

Our results in Fig. \ref{fig:plasmon_OO}(d) and (f) indicate that the background dielectric constant $\varepsilon_r$ plays the same role in controlling the out-of-phase plasmon gap as in the ideal bilayer 2DEG. Specifically, increasing $\varepsilon_r$ reduces the out-of-phase plasmon gap by weakening the interlayer Coulomb interaction, according to Eqs. \eqref{eq:gap_original} and \eqref{eq:gap}. This effect is straightforward, as it is common to bilayers.

\begin{figure}[!htbp]
\centering
\includegraphics[width=0.45\textwidth]{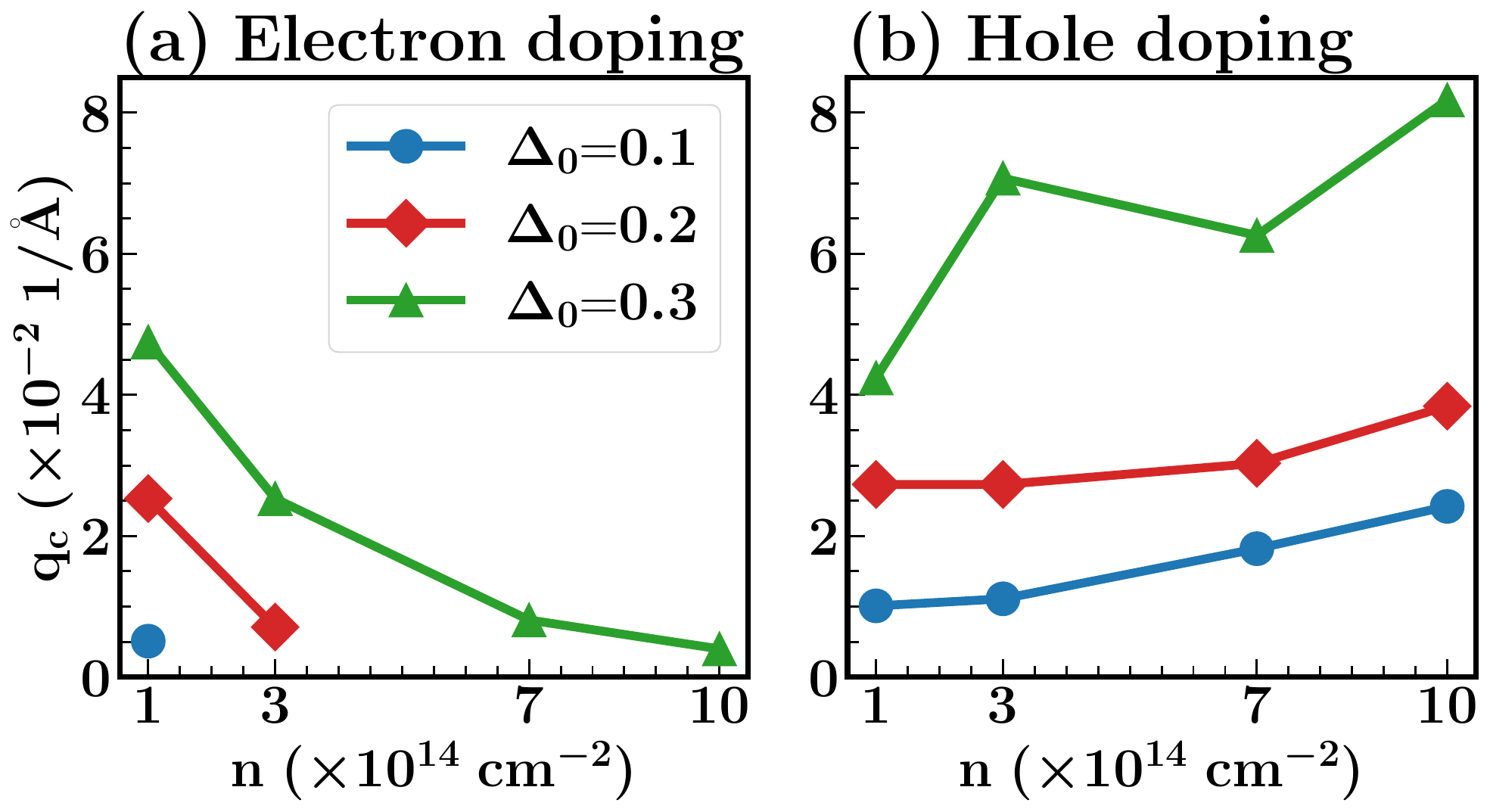}
\caption{The critical momentum of (a) electron- and (b) hole-doped O/O bilayer separating the anomalous and normal dispersions of the out-of-phase plasmon.} 
\label{fig:qc}
\end{figure}

\subsubsection{The anomalous dispersion of the out-of-phase plasmon}

Besides the finite gap, anomalous dispersion is another important feature of the out-of-phase plasmon in a coupled bilayer, directly associated with the EIT. This phenomenon manifests itself as the frequency first decreasing with increasing momentum and then increasing once the momentum exceeds a critical value. For a coupled bilayer 2DEG, the out-of-phase mode has the long-wavelength dispersion\cite{bilayer_2DEG_DasSarma}
\begin{equation}
 \omega_-^2(q)=\Delta^2 + C_1q +C_2q^2,
\end{equation}
with $C_1<0$ and $C_2>0$. The term $C_1=-(\Delta^2-(\Delta_{SAS}^{\rm eff})^2)\frac{h}{2}$ is responsible for the anomalous dispersion and is determined by both the interlayer Coulomb interaction and the EIT. The term $C_2$, which depends in a complicated way on $q_{TF}$, $h$, $n$, and $\Delta_{SAS}^{\rm eff}$, governs the normal dispersion. The competition between these two terms gives rise to a critical momentum
\begin{equation}
q_c=-\frac{C_1}{2C_2}.
\label{eq:qc}
\end{equation}
As shown in Fig. \ref{fig:band_ihs_OO}, anomalous dispersion also occurs in the O/O bilayer. We present the critical momentum for the electron- and hole-doped O/O bilayers in Fig. \ref{fig:qc}.

\begin{figure}[!htbp]
\centering
\includegraphics[width=0.48\textwidth]{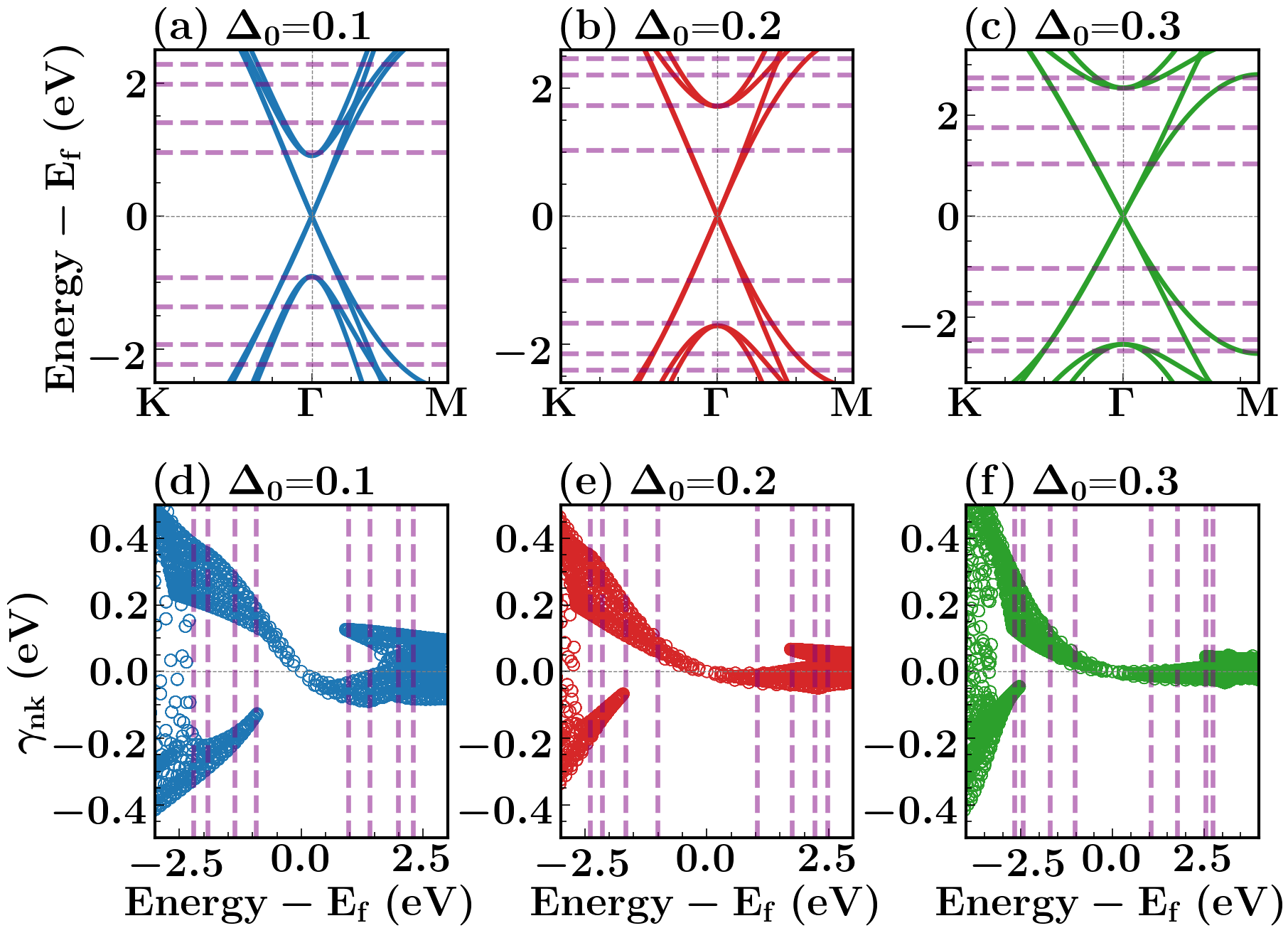}
\caption{Band structures in (a)--(c) and $\gamma_{nk}$ values in (d)--(f) of the G/O bilayer with $\Delta_0=0.1$, $0.2$, and $0.3$. The purple dashed lines indicate the chemical potentials ($\mu$) for four electron and hole concentrations: $10^{14}$, $3\times10^{14}$, $7\times10^{14}$, and $10^{15}\ \mathrm{cm}^{-2}$. Positive and negative values of $\mu$ correspond to electron and hole doping, respectively. A $\mu$ closer to the Fermi energy corresponds to a smaller electron or hole concentration.} 
\label{fig:band_ihs_GO}
\end{figure}

For the electron-doped O/O bilayer at fixed $\Delta_0$, both $\Delta$ and $\Delta_{SAS}^{\rm eff}$ are constant, so $|C_1|$ remains unchanged as the electron concentration varies. Therefore, we expect that increasing the carrier concentration increases $C_2$ and reduces the critical momentum. For a fixed electron concentration, enhancing the Kekul\'{e} distortion increases the critical momentum, which is associated with the increase of $|C_1|$ via the increase of the out-of-phase plasmon gap.

For the hole-doped O/O bilayer, the dependence of the critical momentum on the hole concentration is opposite to that of the electron-doped O/O bilayer, namely, increasing the hole concentration roughly increases the critical momentum. From the results for the electron-doped bilayer, we know that a larger carrier concentration leads to a smaller $q_c$ by increasing $C_2$, but for the hole-doped bilayer it largely increases $|C_1|$ by enhancing the EIT, according to the relation $|C_1|=(\Delta_{SAS}^{\rm eff})^2m^*e^2h^2/\varepsilon_r$ when the bonding and antibonding bands are occupied. Similar to the electron-doped case, $q_c$ increases as the Kekul\'{e} distortion is enhanced, which slightly increases the out-of-phase plasmon gap and reduces the EIT strength, as shown in Fig. \ref{fig:Dsas_hole}. Therefore, the electron-hole asymmetry of the EIT is the main reason for the opposite dependence of the anomalous dispersion on the carrier concentration.

Just as for the out-of-phase plasmon gap, the background dielectric constant $\varepsilon_r$ also straightforwardly controls the anomalous dispersion. Specifically, increasing $\varepsilon_r$ reduces both the $|C_1|$ term and the critical momentum, as shown in Fig. \ref{fig:plasmon_OO}(d) and (f), which is also a common phenomenon in bilayers.

\section{Results and discussion for G/O bilayer}

\begin{figure*}[!htbp]
\centering
\includegraphics[width=0.9\textwidth]{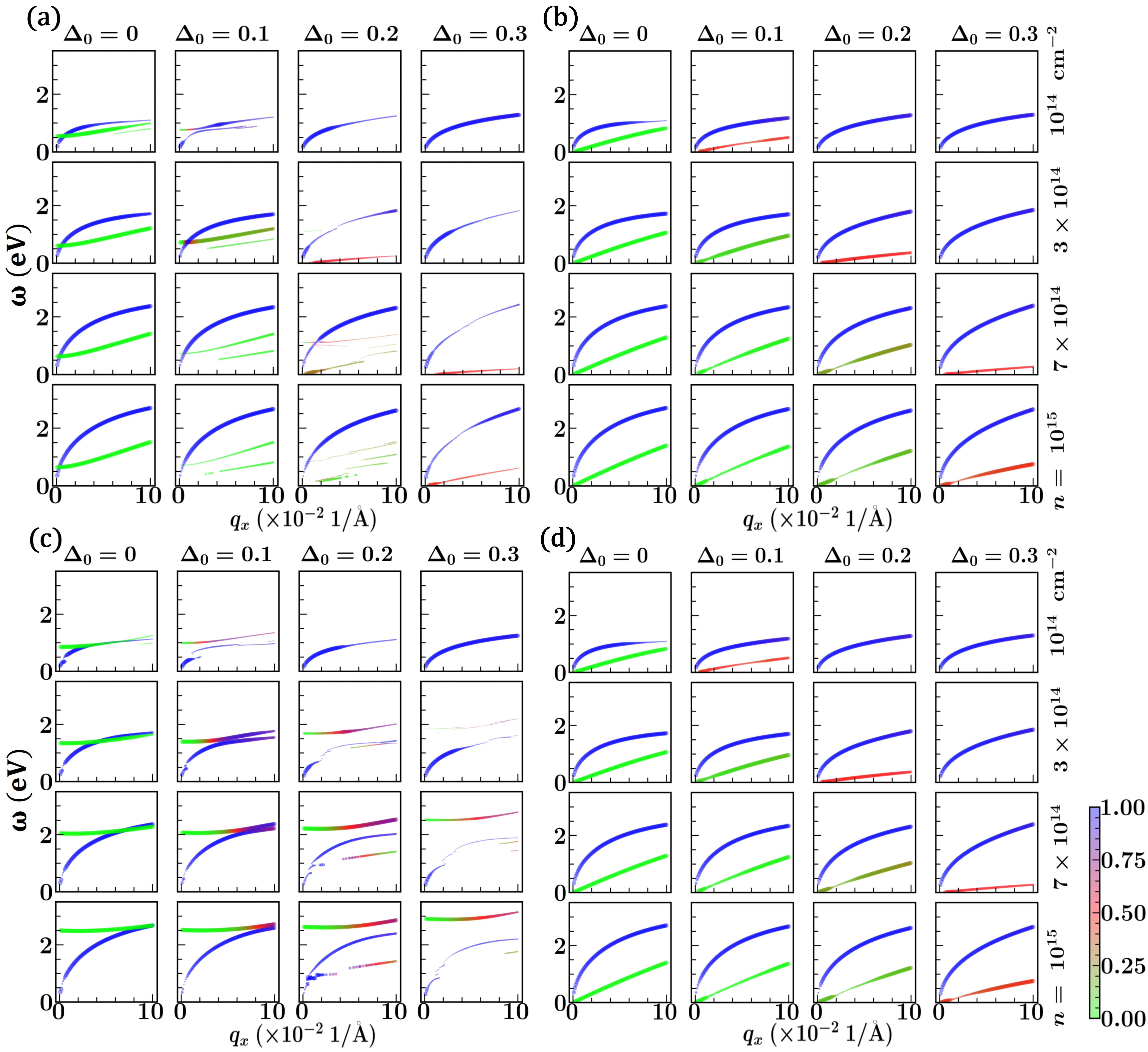}
\caption{Plasmons of the electron-doped [(a) and (b)] and hole-doped [(c) and (d)] G/O bilayer. In (b) and (d), the EIT is manually turned off by removing the interlayer hoppings. In each subplot, from top to bottom, $n$ varies from $10^{14}$ to $10^{15}\ \mathrm{cm}^{-2}$ through $3\times10^{14}$ and $7\times10^{14}\ \mathrm{cm}^{-2}$; from left to right, $\Delta_0$ is $0, 0.1$, $0.2$, and $0.3$. For all calculations in this figure, the background dielectric constant is $\varepsilon_r=1$. The colorbar and dot size have the same meaning as those in Fig. \ref{fig:plasmon_OO}.} 
\label{fig:plasmon_GO}
\end{figure*}

In Fig. \ref{fig:band_ihs_GO}(a)--(c), we present the band structures of the G/O bilayer with $\Delta_0=0.1$, $0.2$, and $0.3$. Owing to the lifted layer degeneracy in the metal/semiconductor heterostructure, the electronic states around the Fermi energy originate mainly from the graphene layer. Enhancing the Kekul\'{e} distortion enlarges both the band gap of the Kek-O layer and the degree of layer-degeneracy breaking. Thus, the Kekul\'{e} distortion strength can tune the EIT even when the chemical potential crosses the bands of both the graphene and Kek-O layers. Because of the strong layer-degeneracy breaking, we cannot directly extract the EIT, but we can still infer it indirectly from the interlayer hybridization, since they exhibit similar physics and energy dependence. In Fig. \ref{fig:band_ihs_GO}(d)--(f), we show the $\gamma_{nk}$ values of the G/O bilayer for several $\Delta_0$ values. The electron-hole asymmetry of the interlayer hybridization persists: the $\gamma_{nk}$ values for the electron-doped bilayer are much smaller than those for the hole-doped bilayer, and increasing $\Delta_0$ further reduces $\gamma_{nk}$ for electron doping.

In Fig. \ref{fig:plasmon_GO}(a) and (c), we show the plasmons for the electron- and hole-doped G/O bilayer. For comparison, the plasmons without the EIT are shown in Fig. \ref{fig:plasmon_GO}(b) and (d), where the EIT is manually turned off by removing the interlayer hoppings. The in-phase plasmons exhibit trivial behavior similar to that in the O/O bilayer and are unaffected by the EIT. Moreover, in the absence of the EIT, the out-of-phase plasmons recover the $\omega \sim q$ dispersion, because the charge oscillation of the bilayer can be regarded as isolated oscillations within each layer coupled by the interlayer Coulomb interaction.

For both electron- and hole-doped G/O bilayers, at the parameter combinations of $n=10^{14}\ \mathrm{cm}^{-2}$ with $\Delta_0=0.2$, and $n=10^{14}$ and $3\times10^{14}\ \mathrm{cm}^{-2}$ with $\Delta_0=0.3$, the chemical potential crosses only the bands of the graphene layer, and the $\gamma_{nk}$ values are small as well, as shown in Fig. \ref{fig:band_ihs_GO}. Consequently, only in-phase plasmons appear, originating solely from charge oscillations within the graphene layer. We now focus on other combinations of carrier concentration and Kekul\'{e} distortion, for which the chemical potential crosses the bands of both the graphene and Kek-O layers. For $\Delta_0=0.1$, the layer-degeneracy breaking is the weakest among the three Kekul\'{e} distortions, and the plasmons of both electron- and hole-doped bilayers retain more features of the bilayer without Kekul\'{e} distortion. However, the broken layer degeneracy still causes the out-of-phase plasmons to become mixed-phase or even in-phase. For $\Delta_0=0.2$ and $0.3$, several new out-of-phase and mixed-phase plasmons appear, but they are strongly Landau damped. Importantly, for the electron-doped G/O bilayer with $\Delta_0=0.3$, the plasmon dispersions in the low-frequency region are identical to those obtained without the EIT, as shown in Fig. \ref{fig:plasmon_GO}(b). Thus, from the perspective of the dispersion, the EIT is effectively turned off by a strong Kekul\'{e} distortion $\Delta_0=0.3$ in the electron-doped G/O bilayer, although it still plays a non-negligible role in enhancing Landau damping compared with the case without the EIT.

\section{Summary}

In this paper, the collective excitations of graphene bilayer with O-type Kekul\'{e} distortion are studied using the tight-binding method and the random phase approximation. Both types of bilayers, namely O/O and G/O, exhibit strong electron-hole asymmetry in the EIT and interlayer hybridization. 

For the O/O bilayer, the EIT is almost constant for electron doping, but depends sensitively on hole concentration. The out-of-phase plasmon in the hole-doped case has a larger gap and stronger anomalous dispersion than that in the electron-doped case. In addition, the plasmons can be effectively tuned by the carrier concentration, the background dielectric constant, and the Kekul\'{e} distortion strength.

For the G/O bilayer, where the electronic states around the Fermi energy originate mainly from the graphene layer, only plasmons corresponding to charge oscillation inside the graphene layer exist at small carrier concentrations. Increasing the carrier concentration can introduce the out-of-phase and mixed-phase plasmon. In addition, for a large Kekul\'{e} distortion of $\Delta_0=0.3$, the electron-doped G/O bilayer has the plasmon dispersions identical to the case with the EIT removed manually, although the EIT still plays non-negligible role in enhancing the Landau damping. 

Unlike the ideal bilayer 2DEG, which has been widely explored in the literature, bilayers of real 2D materials often exhibit electron-hole asymmetry in the interlayer hybridization and also the EIT. Therefore, we expect that electron-hole-asymmetry-associated plasmons exist widely. Thus, switching the doping type between electron and hole should be a possible method to tune the performance of plasmon-based electronic and optical devices.

\begin{acknowledgments}
This work was supported by the Science and Technology Development Plan Project of Jilin Province, China (Grant No. 20240101323JC), and the Fundamental Research Funds for the Central Universities (Grant No. 2412025QD011). 

\end{acknowledgments}

\bibliography{ref}
\end{document}